# Molecular Packing Motifs Determine Charge-Transfer and Carrier Dynamics in Molecular Heterosystems: the Case of Pentacene - Perfluoropentacene

*Andre Rinn§, Tobias Breuer§, Julia Wiegand†, Michael Beck†, Jens Hübner†, Michael Oestreich†, Wolfram Heimbrodt§, Gregor Witte§, and Sangam Chatterjee\*§$*

§Faculty of Physics and Materials Sciences Centre, Philipps-Universität Marburg, Renthof 5, D-35032 Marburg, Germany

†Institut für Festkörperphysik, Leibniz-Universität Hannover, Appelstrasse 2, D-30167 Hannover, Germany

$Institute of Experimental Physics I, Justus-Liebig-University Giessen, Heinrich-Buff-Ring 16, D-35392 Giessen, Germany

\*E-mail: sangam.chatterjee@physik.uni-giessen.de

**Abstract**

The great majority of electronic and optoelectronic devices depends on interfaces between n-type and p-type semiconductors. Finding such matching donor-acceptor systems in molecular crystals remains a challenging endeavor. Structurally compatible molecules may not necessarily be suitable with respect to their optical and electronic properties: large exciton binding energies may favor bound electron-hole pairs rather than charge separation by exciton dissociation, and free, band-like transport is challenging to achieve as hopping commonly dominates charge motion. Structurally well-defined pentacene-perfluoropentacene heterostructures in different polymorphs and molecular orientations are model systems to study the relation of packing motif and optical properties. These heterosystems feature two characteristic interface-specific luminescence channels at around 1.4 and 1.5 eV. Their relative emission strength strongly depends on the molecular alignment of the respective donor and acceptor molecules. Evaluating their dynamics in comparison with the corresponding unitary films reveals the role of singlet-triplet intersystem crossing and different channels for carrier injection into the interface-specific resonances.

Semiconductor p-n junctions are well established for inorganic semiconductor heterostructures through doping the constituents with great versatility and enormous precision. Creating matching donor-acceptor molecule pairs in molecular semiconductors with suitable electronic properties for the respective device is demanding. Hence, many efforts of the scientific community are targeted towards identifying and characterizing suitable organic donor-acceptor pairs.[1, 2, 3] The electronic properties play a leading role for the functionality of a material system as a p-n junction. For possible applications in photovoltaics, the absorption cross section of both molecules is essential for high carrier yields from irradiation by sunlight. Multiple different chromophores are carefully designed and evaluated for their applications in donor-acceptor junctions with promising results, see, *e.g.*, 1, 4.

Both, electronic and structural properties need to match for material systems suitable for optoelectronic devices. Neither may structurally compatible molecules be necessarily suitable with regard to their electronic properties nor may electronically suitable pairs be structurally compatible. The most fundamental aspects regarding the formation of donor-acceptor pairs are presumably the electron affinities of both materials. They define the donor and acceptor levels at their interface. Baring stronger intermolecular coupling, the interaction by quadrupole moments plays an important role for such donor acceptor interaction[5] and, in particular, for the stacking of the molecules at an interface.[6] The existence of possible interface charge-transfer excitons whose electrons (holes) are located in the acceptor (donor) material depends on the exact level alignment of both molecules HOMO and LUMO orbitals. In general, exciton binding energies in molecular solids are large. This inhibits exciton dissociation and favors charge transport by exciton hopping over band like transport usually observed in inorganic solids.[7,8]

Careful sample design is often able to solve this challenge. For example, it can lead to π-π-stacking in acene-type organic solids. Such a molecular packing, while repulsive in nature, can be achieved by choosing appropriate growth parameters.[9] It increases the coupling strength by overlapping the delocalized π-electron systems of two aromatic molecules in such solids. Due to the anisotropy of many molecular systems, the structural compatibility of two molecules not only depends on the lattice parameters of the crystals, but also on the crystalline polymorphs and molecular orientations within the unit cells. This structural intricacy of molecular solids adds an additional degree of complexity to electronic considerations. Most materials exist in multiple polymorphs. Even minute structural changes may result in notably different optoelectronic properties.[10] Pentacene (PEN), for example, features several bulk realizations.[11,12] Singlet exciton fission has been observed in PEN, i.e., one singlet-type exciton ($S_1$) splits non-radiatively into two triplet-type excitons ($T_1$), thereby effectively doubling the number of charges available for current generation. The singlet exciton fission process can only occur for $2\,E_{T1} \leq E_{S1}$ without significant phonon assistance. Organic photovoltaics thus can conceptually achieve efficiencies beyond the Shockley-Queisser limit.[13] Singlet exciton fission influences the emission spectra of those materials since it non-radiatively depopulates the bright S1 exciton and fast carrier decay as well as low luminescence efficiencies result.

Slip-stacked crystalline structures of aromatic molecules generally enhance singlet exciton fission as this process involves two states on neighboring molecules with significant wave-function overlap.[14,15] Fission is generally believed to be mediated by a more delocalized excimer-like states in the solid state; these are sometimes referred-to as charge-transfer states in the literature,[16,17] inferring a certain degree of delocalization across more than one molecule. These are not to be confused with states across an internal interface with charge-transfer character, which we will refer to as charge-transfer (CT) states in this work. The latter exhibit a spatial localization of holes and electrons in the donor and acceptor molecules, respectively, thereby, forming a charge-transfer exciton across the internal donor-acceptor interface. Besides singlet exciton fission, triplet exciton fusion can be observed in organic solids. This back transfer from the triplet to the singlet system is significant for increasing the efficiency of fluorescent organic light-emitting diodes[18,19] and is also responsible for the delayed fluorescence feature observed in time-resolved photoluminescence (TRPL) measurements as long-living luminescence tail. Triplet exciton fusion can only occur for $2E_{T1} \geq E_{S1}$. For the materials presented in this paper, the available data on the triplet and singlet exciton systems suggest that both singlet exciton fission and triplet exciton fusion can occur simultaneously, further increasing the degree of complexity for the interpretation of optical measurements.

Prototypical charge-transfer systems displaying strong singlet exciton fission are PEN-fullerene heterostructures.[20] They have been subject to many experimental and theoretical studies, *e.g.*, in Ref. 21,22. In particular, the formation of CT states is in direct competition to singlet exciton fission.[23] However, the structural compatibility of the two molecules is rather limited and renders detailed microscopic studies of the relation between structural and electronic properties as challenging. More promising in this respect are PEN-perfluoropentacene (PFP) heterostructures. Here, the large electronegativity of fluorine turns PFP into a strong acceptor molecule by lowering its electron affinity.[2] PEN and PFP exhibit inverted quadrupole moments, which increases the mutual intermolecular interaction of both molecules.[24] Clear signatures of CT states are observed in intermixed thin films in PL and absorption experiments.[25,26] Intermixed, 1:1 cocrystallized PEN-PFP films, i.e., containing same amounts of PEN and PFP, are stable.[27] This is attributed to an alternating cofacial stacking[28] as found for naphthalene-perfluoronaphthalene intermixed heterostructures.[29] While the PL spectra simply display the individual signals of PEN and PFP at room temperature, those disappear at cryogenic temperatures. Instead, a novel feature is observed at 1.4 eV,[25] which is attributed to a CT excitation between PEN and PFP molecules. Such bulk blends with intermixed heterojunctions can be disadvantageous for application in devices as carrier extraction may potentially be challenging, therefore well-ordered, layered heterostructures may be better suited. Such well-ordered structures of different polymorphs and molecular orientations can be grown with excellent quality well beyond the first few molecular layers by carefully templating epitaxial layers with suitable substrates.[30]

In this work, we present the carrier dynamics of the intermixed and layered heterojunctions of PEN and PFP and study the influence of different packing motifs at the interface. The results are compared to corresponding measurements on the individual constituents forming the heterostructures to identify newly emerging charge-transfer states. TRPL measurements provide information on the energy transfer to CT states as distinct, interface-specific effects. Specifically, we observe an enhancement of the CT signal strength for π-π stacking at the PEN-PFP interface.

**Results and Discussion**

We investigate three different packing configurations of PEN-PFP heterosystems: a bulk heterostructure, i.e., an intermixed 1:1 blend as well as layered heterostructures (heterostacks) with upright molecular orientation and lying molecules orientation with regard to the substrate, respectively. For easier reading, we will refer to structures with upright (lying) molecular orientation with regard to the substrate as standing (lying). All pure-phase constituents of the heterostructures are studied for reference. The latter spectra for the different materials are presented in Figure 1. Panels A and B display the molecular structure of the standing and the lying PEN films, respectively. The standing (lying) films adopt the Campbell (Siegrist)-phase. Both feature a herringbone (HB) packing-motif, however, each with slightly modified lattice parameters and a unique molecular orientation with respect to the substrate.[9] Those differences reflected themselves in the corresponding PL spectra (Figure 1 G). PEN shows two major peaks in both molecular orientations: the free exciton line at around 1.8 eV and the self-trapped exciton line at 1.645 eV. [31] For the free exciton line, the PL from the Siegrist phase is red-shifted by about 50 meV compared to its Campbell phase counterpart. The self-trapped exciton shows no significant spectral shift for the different polymorphs.

The corresponding polymorphs in PFP are depicted in panels C and D. The standing PFP (C) adopts a HB packing motif with the molecules virtually perpendicular to the substrate.[32] The lying film (D) displays a brick-wall structure where the molecules lie flat and parallel to the substrate.[9] In the following, we will refer to the latter as π-stacked polymorph (PSP). Significant differences in the optical properties are found for PFP: the PSP dominantly emits at 1.645 eV while the HB phase luminesces at 1.71 eV. This energy difference of 65 meV in emission is significantly larger than that in absorption, where the separation in energy between the two lowest-lying singlet-excitons measured at 10 K at 1.67 eV (PSP, Supporting Information, Figure S1) and 1.72 eV (HB) is 45 meV. The larger Stokes shift of 25 meV in the PSP between the absorption and emission when compared to the 10 meV of the HB structure suggests a more complex relaxation scheme than mere emission from free excitons. Additionally, the PL from the PSP is more intense by over an order of magnitude for comparable excitation conditions. This may be understood when considering that efficient singlet exciton fission typically quenches the PL. In previous work, singlet exciton fission has been reported for the HB polymorph of PFP.[33] The PSP, however, displays an even shorter π-stacking distance (3.07 Å in PSP, 3.17 Å in HB) as well as a more favorable slip stacking of the molecules.[9, 34] Yet, the PSP shows a much brighter singlet-type emission. These seemingly contradictory observations mandate information about the respective triplet systems. Unfortunately, the energetics of the $T_1$ state has not been reported for the PSP to date, neither experimentally nor theoretically. We hence can only estimate it to be in the vicinity of its HB counterpart. This is supported by the small shift of only 45 meV of the S1 in the PSP when compared to the HB system. However, experimental values for the lowest lying triplet exciton in PFP are not reported to our knowledge. Theoretical estimates have been performed for the HB polymorph, yielding values between 0.88 eV [26] and 0.62 eV.[35] Both energy values are small enough to allow singlet exciton fission within the typical computational accuracy of such calculations. Then, singlet exciton fission should be allowed by energy considerations and even be favored by the crystalline structure in the PSP. However, the bright luminescence that is observed from the PSP of PFP seemingly contradicts these considerations. The discrepancy may be lifted by taking into account triplet exciton fusion. Energetically, $2E_{T1} \sim E_{S1}$ holds true for PFP if we use the $E_{T1}$ value of 0.88 eV, so that fission and fusion could occur simultaneously. This would lead to delayed fluorescence, a long-living PL tail that should be visible in TRPL.

Thus, evaluating the TRPL measurements provides further insight into the carrier dynamics and the nature of the PL emission (Figure 1F). The dominant emission from both PFP polymorphs at 1.71 eV (HB, dark green) and 1.645 eV (PSP, light green) display fast carrier decay with a time constant of around 20 ps. A slightly faster decay of 15 ps is observed in the PSP, which is consistent with its structural properties being well suitable for fast singlet fission. Due to its smaller π-stacking distance, excimer-like states may more easily form in the PSP. Those states are not only a precursor for singlet fission, but could also lower the emission energy. This corroborates the Stokes shift of 25 meV between the PSP absorption and emission.

Longer carrier lifetimes are observed in the self-trapped exciton line of the Siegrist (grey) and the Campbell phase (black) PEN. We find a single-exponential decay with a decay time of 3 ns for the standing structure and 800 ps for the lying one after 150 ps. The absence of long-lived PL signals in the PFP systems rules out delayed fluorescence by triplet exciton fusion. Therefore, the lower triplet energy of 0.62 eV seems somewhat more probable since it would not energetically allow for triplet fusion. The difference in PL intensity of the two PFP structures is tentatively related to the difference in absorption strength at the pump wavelength of 400 nm (Supporting Information, Figure S2). The higher absorption of lying molecules in all PFP polymorphs at higher wavelengths has to be kept in mind for light harvesting applications in organic solar cells. The stark contrast in carrier dynamics between the PFP luminescence and self-trapped exciton emission in PEN can be used as a tool to distinguish the spectrally overlapping PEN and PFP signals in the heterostacks.

Their structural configurations are illustrated in Figure 2 in panels A, B, and C for the intermixture as well as the lying and the standing heterostacks, respectively. The corresponding PL spectra at liquid-helium temperatures are provided in Figure 2D, together with the absorption spectrum of the intermixed-film. The intermixed heterosystem (pink) lacks the monomer PL of pure PEN and pure PFP between 1.8 eV and 1.645 eV. Both disappear with decreasing temperatures while the characteristic CT signal at 1.4 eV increases.[25] From this observation, we conclude that the molecules in the intermixed blend couple to the CT state with a near-100 % efficiency for low temperatures. Virtually all injected carriers are transferred to the CT state before any radiative decay from the bulk states is detectable. A satellite peak is observed at 1.55 eV, which cannot be linked to any monomeric PL signal visible in the pure phases. This may be assigned to a vibronic replica. Alternatively, both peaks could be related to different kinds of CT excitons formed by different molecule pairs with varied mutual molecular alignments.

Similar CT-related signals are visible for the heterostacks. The main PL peak is, however, slightly shifted to 1.35 and 1.37 eV for the standing and lying heterostack, respectively. A significant difference can be seen in the PL intensity. The lying film shows more than twice the intensity of its standing counterpart. This can be linked to the electronic coupling of the aromatic PEN and PFP molecules. For the lying heterostack, the molecules can overlap their π-electron systems at the interface. This increases intermolecular coupling and therefore the formation of CT states. Additionally, more light is absorbed in the pure lying PEN and PFP, contributing to the high signal strength. The satellite peak is found at around 1.49 eV for both heterostacks. Note that the number of molecules neighboring molecules of the respective other kind in the heterostacks is only a fraction of those in the intermixture, even if we take intermixture at the interface of the heterostacks due to intercalations during the growth process into account.[30] Even when considering the enhanced absorption of the lying PFP compared to the standing PFP, this renders the relative brightness of the CT state specifically intriguing.

Pronounced monomer PL of PEN and PFP is found at around 1.8 and 1.645 eV for both heterostacks. These features are completely absent in the intermixed heterostructure. This is congruent with absorption measurements, where the lowest lying exciton peaks of pure PEN and PFP are not visible for intermixed films (Figure 2B).[26] The signals at around 1.8 eV are attributed to the free exciton PL of PEN by comparing them to the emission of corresponding pure-phase PEN PL. The PL of the lying heterostack is redshifted with regard to the signal from the standing heterostack, similarly to the lying and standing pure PEN. The interpretation of the PL line at 1.645 eV in the heterostacks is more intricate: comparing Figures 1C and 2D reveals a spectral overlap of the emission from the self-trapped PEN exciton and PFP exciton with the PL at 1.645 eV in the lying heterostack. Hence, an assignment of this emission peak in the heterostacks requires closer evaluation. For the standing heterostack, the data suggest a dominating contribution of the PEN-related signature. The PL from the self-trapped PEN exciton is much more intense than the PFP emission in the pure phases. However, the corresponding, heterostack shows no emission related to the standing PFP (at 1.71 eV). This suggests that the excitations formed in the PFP layer are transferred to the molecular interface where they relax into the CT state and recombine radiatively. More evidence for this interpretation can be found in time-resolved measurement. Unfortunately, such a clear assignment to an individual contribution of PEN cannot be made for the lying heterostack. Both, the PEN and PFP emission show significant emission strength at the exact same energy. Hence, the PL signal at 1.645 eV cannot be directly linked to any of the individual constituting of the heterostack. Regardless of its origin, this PL line lies in close proximity of the absorption peak of the CT state at 1.6 eV for both heterostacks, as shown in Fig 2D. In this context, reabsorption or Förster resonance energy transfer from the bulk materials into the CT state appears possible. These effects should be more pronounced in the lying heterostack, due to the high relative brightness of the PL line at 1.645 eV. This can contribute to the more pronounced CT signal visible for the lying heterostack (Figure 2D). All PL and TRPL measurements shown in this study are excited with an excitation wavelength of 400 nm. Therefore, it is reasonable to assume that most carriers are, therefore, excited in the energetically higher PEN and PFP levels before they may relax into the CT state visible in the heterosystems. The existence of an absorption line for the CT state in the intermixed heterosystem suggests that a direct excitation of the CT state should be possible. Indeed, excitation with a laser wavelength of 760 nm could reproduce the PL of the CT state for all three heterosystems without any noticeable spectral differences.

An energy diagram of the relevant states is given in Figure 3. The alignment of the HOMO (-4.9 eV for PEN, -6.65 eV for PFP) and the LUMO (-2.7 eV for PEN, -4.55 eV for PFP) levels of the standing PEN and PFP crystals are given in black and green, respectively.[38] The first $S_1$ and T1 (0.86 eV [36] and 0.62 eV [35]) exciton states are included for both materials at positions in nergey below the LUMO level. Their difference corresponds to the respective exciton binding energies. To showcase its proximity in energy to the $S_1$ state, an energy level for twice the triplet energy ($2E_{T1}$) is added. For clarity, all intermediate states in singlet exciton fission and triplet exciton fusion like excimer CT states or correlated triplet pair states are omitted. Interface states are often described as a correlated exciton state formed by a hole in the donor's HOMO level and electron in the acceptor's LUMO level. However, the luminescence line attributed to the interface CT exciton is observed at ~1.4 eV for all three heterosystems. The CT transition in intermixed films is located at 1.6 eV in absorption measurements (Figure 2D). This does not correspond to any conceivable intersystem transition depicted in Figure 3 on the basis of the used values for the frontier orbital energies.[37] This suggests that a new state is formed at the interface at least 1.6 eV above the PEN or PFP HOMO level in energy. This pure interface state is depicted in blue in Figure 3.

Next, we turn to the emission dynamics of the heterosystems. The time-resolved data of the intermixture are presented in Figure 4A for three distinct spectral regions: the pure PEN channel at around 1.8 eV, the PFP and self-trapped PEN channel at 1.645 eV, and the CT channel at around 1.4 eV. The dynamics of the PEN free exciton channel in the two heterostacks are provided in Figure 4B along with the dynamics of the corresponding pure phases on a logarithmic scale. A comparison between these datasets suggest the dynamics of the PEN free exciton emission to be rather unaffected by the existence of the PEN-PFP interface (Figure 4B). All transients show a PL decay at the limit of our time resolution. Such fast dynamics of a few picoseconds or below are consistent with previously reported singlet exciton fission times in PEN.[20, 38]

We proceed with the examination of the PL emission at 1.645 eV. Significant differences between the transients of the pure phases and the heterosystems are observed, *cf.* Figure 4C, where the transient PL is compared to the emission from the pure phases at the same energy. For an adequate comparison, we have to add up the transients for the self-trapped PEN and lying PFP exciton before normalization and compare these combined transients to the observed signal in the heterosystem. This combination is not necessary for the standing structures since the dynamics of the standing heterostack are well reproduced by only the self-trapped exciton line of standing PEN, while the corresponding PFP line observed at 1.71 eV in the pure phase was not observed in the standing heterostack. Thus, we compare the combined-transient of the pure-phase lying PFP and PEN with the signal observed in the heterostack. The transient PL of the heterostack shows a steep drop in the first 200 ps, followed by a long-lived luminescence tail. One can easily conclude that the spectrum is dominated by the PFP PL during the first 200 ps and by the emission from the self-trapped exciton in PEN thereafter when taking into account the PL decay dynamics of pristine PEN and PFP. Plotting the sum transient of both constituents (deep purple) excellently replicates the decay times at short and long times. The sum transient and the corresponding signal from the lying heterostack only differ in relative weight of the PEN and PFP contribution. The data are in good agreement considering any potential variation in thickness of the layers in pristine films and heterostructures.

These findings assist in interpreting the CT state dynamics displayed in Figure 4D. The carrier dynamics are in general found to be slow when compared to the free exciton decay of PEN and the observed PFP lines. Similar, bi-exponential decay dynamics are observed for all three material systems with carrier lifetimes of around 50 ps for the faster decay channels and 800 ps for the slower ones. The latter render singlet exciton fission in the CT system very unlikely, in stark contrast to the individual constituents of the heterostructures. Disregarding the intensities, the heterostacks of both orientations display almost identical carrier dynamics. Differences are observed only for the intermixed phase: the initial, faster decay channel is less pronounced for the intermixed PEN-PFP. All transients show identical decay times after roughly 400 ps. Thus, the intrinsic carrier dynamics of the CT state in all PEN-PFP heterostructures are barely influenced by the long-range order of the material. However, differences in the molecular alignment between the constituents of the heterostructure influence the initial carrier decay. This can be attributed to defect states at the interface. The faster decay in the heterostacks indicates a good and comparatively defect-free intermixing of both constituents in the bulk heterostructure. This is important to keep in mind for any potential device application.

**Conclusions**

We investigate the charge and energy transfer characteristics of the molecular donor-acceptor pair PEN-PFP. Comparing the PL of pure PEN and PFP for different packing motifs with the corresponding heterosystems, intrinsic bulk signatures are clearly distinguished from interface-related features. A strong energetic shift and significant suppression of the PFP PL are found in the standing HB structure when compared to its π-stacked counterpart. Possible links to singlet exciton fission and triplet exciton fusion are investigated: fast carrier dynamics and a noticeable Stokes shift between absorption and emission in the PFP PSP could be due to the efficient formation of excimer-like states, precursors for singlet exciton fission. The difference in PL intensity is assigned to the difference in absorbance for lying and standing PFP. Furthermore, we observe CT exciton emission in all three types of heterostructures. The CT states apparently form directly at the interface with great efficiency. The observed decay times of these states are an order of magnitude longer than those observed in the pure materials. Furthermore, π-stacking at the interface seems to favor the formation of CT states as inferred from the relative intensities of this transition for all three types of heterostructures. An increase of the CT emission strength is observed for the lying heterostack due to better coupling between the stacks constituents and better light harvesting for lying molecular orientation when compared to the standing heterostack. The pure-phase signatures in the heterostacks suggest an energy transfer into the CT state for the lying heterostack by Förster resonance energy transfer or by reabsorption, further explaining the high PL intensities for this material. The carrier dynamics of the CT state display similar decay in both heterostacks; however, differences are found for the first few hundred picoseconds after excitation when compared to the intermixture: the latter shows a less-pronounced initial quench. This is assigned to a higher concentration of available trap states in the heterostacks compared to the blend. Overall, the insight into the carrier dynamics in organic heterostructures gained through these results may serve as a basis for designing future advanced materials systems.

**Materials & Methods**

**Sample preparation**

The PEN (Sigma-Aldrich) and PFP (Kanto Denka Kogyo Co.) samples are grown under high-vacuum conditions by molecular-beam deposition from resistively heated Knudsen cells. Upright and lying molecular configurations are achieved by using natively oxidized Si-wafers and graphene-coated quartz substrates, respectively, as detailed in ref. 28. All film structures are verified by X-ray diffraction analyses. The molecular flux during film growth is monitored by a quartz crystal microbalance and typically balanced at 6 Å/min. The films are prepared with nominal thicknesses of 20 nm for the unitary films and 40 nm for the heterostructures (in each case with stoichiometric mixture of both constituents). The substrates are kept at room temperature during film preparation to minimize molecular intermixture at the interface for the layered heterostructures.[30]

**Time-resolved photoluminescence**

The TRPL in the visible regime is acquired using a streak camera setup with all-reflective optics.[39] A Ti:sapphire laser emitting 100-fs pulses at a repetition rate of 78 MHz is frequency-doubled for excitation. All samples are excited with 400 or 428 nm light at 10 mW of power into a 3µm diameter spot using a confocal beam path used for excitation and detection. The PL signal is spectrally dispersed by a grating spectrograph; the spectra are acquired by a cooled CCD camera, while time-resolved data for all pure-phase PL channels is recorded by a streak camera equipped with a S20 cathode yielding a time resolution of <2ps. For the TRPL in the near infrared regime all samples are excited using a frequency-doubled Ti:sapphire laser at a wavelength of 405 nm with a repetition rate of 80 MHz and a pulse width of 200 fs. The laser is focused onto the sample with a spot diameter of 20 µm and a power of 30 mW. The PL signal is dispersed by a grating spectrograph and the temporal decay of the charge-transfer PL is recorded with a streak camera (Hamamatsu C5680-27 with cooled CCD) in photon counting mode with a time resolution of 8 ps.

**Figure 1**

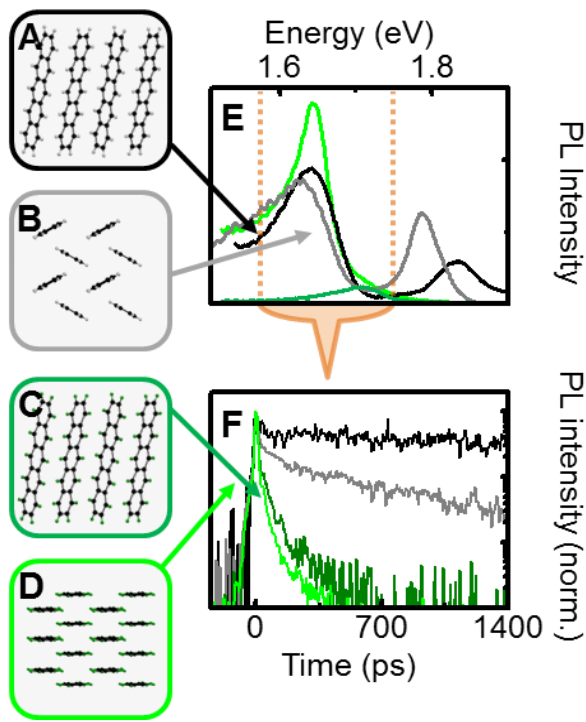

**Figure 1.** Schematic depiction of the crystalline structure for all pure PEN and PFP structures investigated in different molecular orientation (A, B, C, and D) and the corresponding PL spectra at 6 K. (E). The corresponding decay dynamics are given in panel F; the self-trapped exciton PL is shown in the case of PEN.

**Figure 2**

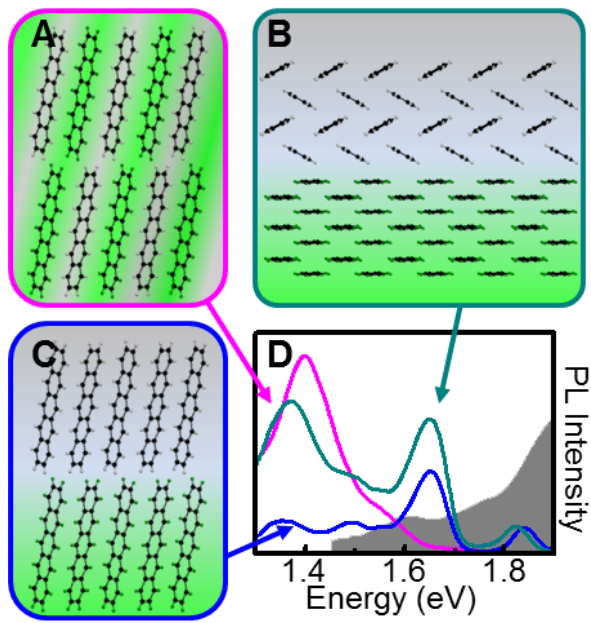

**Figure 2.** Schematic crystalline structures of the investigated PEN-PFP heterosystems: A the intermixed bulk phase, B the lying heterostack, and C the standing heterostack. The corresponding TRPL data are given in D; the linear absorption of the bulk heterostructure is given for reference (grey)

**Figure 3**

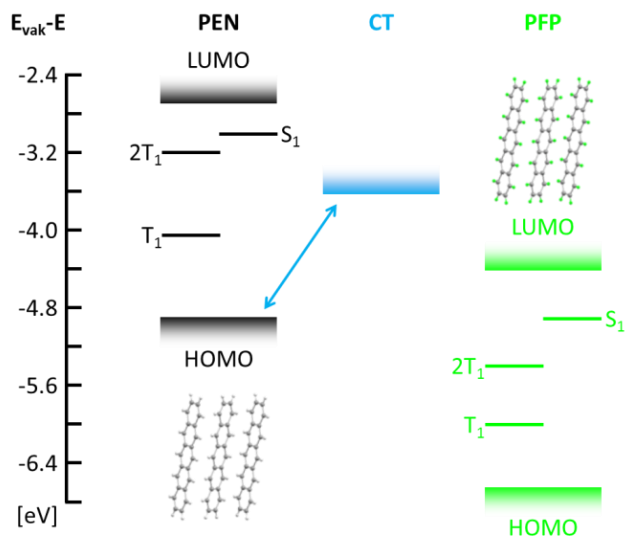

**Figure 3.** Energy alignment of the HOMO and LUMO levels of layered PEN and PFP in upright configuration at the interface as derived from Refs. 26, 35, 37, 38, 40. The lowest singlet and triplet excitons are included below the LUMO level corresponding to their exciton binding energy. The energy of twice the triplet excitation is given to highlight its relative position with regard to the lowest singlet state. The PEN HOMO→PFP LUMO transition should lie at 0.4 eV on the basis of the available frontier orbital energies, which is far below the observed 1.6 eV for the CT absorption signal (Figure 2D). Therefore, we propose the existence of completely new interface states ~1.6 eV above the HOMO levels of one or both materials (blue).

**Figure 4**

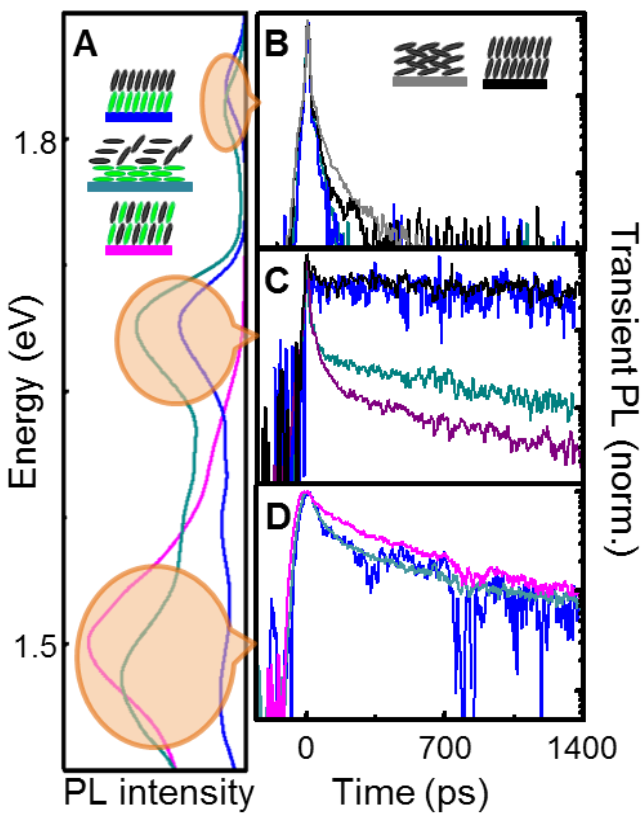

**Figure 4.** TRPL data of PEN-PFP heterosystems in comparison to the corresponding transient PL from the pure phases. The individual energy ranges of interest are highlighted in A. The peak around 1.8 eV (B) is assigned to PEN. The region around 1.645 eV (C) cannot be uniquely assigned to either layer for the lying heterostack as pure lying PEN and PFP contribute. The data are compared to the sum (deep purple) of the PEN and PFP transients at ~1.645 eV. The dynamics of the CT state are shown in D.

*Corresponding Author*

[*] E-mail: sangam.chatterjee@physik.uni-marburg.de

*Author Contribution.* TB and GW prepared the samples and provided the structural characterization. JW, MB, JH and MO measured the TRPL of all three CT states in the near infrared. AR, WH, and SC performed all other experiments and analyzed the data. The manuscript was written through contributions of all authors. All authors have given approval to the final version of the manuscript.

*Acknowledgment.* Financial support was provided through SFB 1083 "structure and dynamics of internal interfaces".

*Conflict of Interest.* The authors declare no competing financial interest.

*Abbreviation.* PEN, pentacene; CT, charge transfer; HB, herringbone; PFP, perfluoropentacene; HOMO, highest occupied molecular orbital; LUMO, lowest unoccupied molecular orbital; TRPL, time-resolved photoluminescence.

*Supporting Information:* The supporting information contains linear absorption spectra of the PSP PFP polymorph in the visible and ultra violet range. Additionally, all observed PL decay times are listed. This material is available free of charge *via* the Internet at http://pubs.acs.org.